**Anomalous temperature evolution of the electronic structure of FeSe**

Y. Kushnirenko [1], A. A. Kordyuk [2,3], A. Fedorov [1,4], E. Haubold [1], T. Wolf [5], B. Büchner[1] and S. V. Borisenko [1]

1. IFW Dresden, P.O. Box 270116, D-01171 Dresden, Germany
2. IMP of National Academy of Sciences of Ukraine, 03142 Kyiv, Ukraine
3. Kyiv Academic University, 03142 Kyiv, Ukraine
4. II. Physikalisches Institut, Universität zu Köln, Zülpicher Strasse 77, 50937 Köln, Germany
5. Institute for Solid State Physics, Karlsruhe Institute of Technology, 76131 Karlsruhe, Germany

**We present ARPES data taken from the structurally simplest representative of iron-based superconductors, FeSe, in a wide temperature range. Apart from the variations related to the nematic transition, we detect very pronounced shifts of the dispersions on the scale of hundreds of kelvins. Remarkably, upon warming the sample up, the band structure has a tendency to relax to the one predicted by conventional band structure calculations, right opposite to what is intuitively expected. Our findings shed light on the origin of the dominant interaction shaping the electronic states responsible for high-temperature superconductivity in iron-based materials.**

Iron-based superconductors (IBS) continue to represent another class of materials with unknown mechanism of pairing at high temperatures. Electronic structure of iron pnictides and chalcogenides has two essential deviations from the predictions of conventional band structure calculations and these deviations may hold the key to understand the phenomenon. First pronounced departure from LDA calculations is the strong renormalization of the valence band with orbital dependent factors ranging from 2 to 9 [1 - 4]. This behavior has been successfully explained in the framework of DMFT calculations by considering the significant exchange interaction J [5 - 7]. The second robust and generic for all IBS families experimental fact is the so called "blue/red shifts" which result in mutually opposite energy shifts of the dispersions near the center and the corner of the Brillouin zone (BZ) [2, 8, 9, 10]. Such shifts lead, in particular, to the shrinking of the Fermi surfaces (FS) in comparison with the calculated ones and bring the van Hove singularities closer to the Fermi level [11]. There are several theoretical approaches to explain such shifts [12 - 19], but neither the consensus nor the quantitative agreement with the experiment is reached.

In this Letter we report an unusual temperature dependence of the low-energy electronic structure in FeSe. The energy location of the electronic dispersions clearly changes with temperature and these variations are momentum dependent. The blue/red shifts tend to disappear with temperature and Fermi surfaces grow in size thus bringing the electronic structure closer to the calculated one. We consider several scenarios which can explain the observed anomaly.

ARPES data have been collected at I05 beamline of Diamond Light Source [20]. Single-crystal samples were cleaved in situ in a vacuum better than $2 \times 10^{-10}$ mbar and measured at temperatures ranging from 5.7 to 270 K. Measurements were performed using linearly polarized synchrotron light, utilizing Scienta R4000 hemispherical electron energy analyzer with an angular resolution of 0.2° – 0.5° and an energy resolution of 3 meV. Samples were grown by the KCl/AlCl3 chemical vapor transport method.

We start presenting the ARPES data by showing the typical Fermi surface map of FeSe in Fig. 1a. There are several sheets: one hole-like located in the center of the BZ and two electron-like located at the corners. The shape of all FSs is modified by the electronic nematicity below 90 K and by presence of the domains [2, 9, 17, 21 - 28]. The circular pocket at the center in the tetragonal phase (shown schematically in Fig.1 a) is replaced by two slightly elliptical pockets from different domains and the crossed ellipses in the corners are more elongated [29]. As has been pointed out in previous ARPES reports, the experimental Fermi surfaces are noticeably smaller than those obtained by the band structure calculations [1, 2, 9, 21 - 28]. Panel (b) of

Fig. 1 clarifies why this is the case. It shows the experimental dispersions along the diagonal direction of the BZ running through both discussed regions of the k-space. It is seen that both, hole- and electron-like bands have their extrema close to the Fermi level and this makes the corresponding Fermi surfaces small.

We have recorded the temperature evolution of these two main constructs from 6 to 270 K. The results are shown in Fig. 2. It is important to distinguish the modifications due to nematic transition occurring at ~90K from global temperature-induced changes on a larger temperature scale. As was shown earlier [2], orthorhombicity causes inequality of the dispersions along ΓX and ΓY directions and results in the small splitting of the features in the ARPES data collected from overlapping domains. The splitting starts to occur at around 90K and relatively quickly reaches its maximum in a manner, typical for an order parameter. We do not focus on the details of this effect here and concentrate instead on another trend, also clearly seen from the spectra. Already a visual inspection of Fig. 2 clearly implies the monotonic shifts of the features with temperature: electron-like dispersions from the corner of the BZ in the upper row of panels move downward to higher binding energies while the hole-like dispersions in the lower row move upwards.

Further detailed quantification of this temperature evolution is presented in Fig. 3. Here we show intensity plots for single energy distribution curves (EDC) divided by the Fermi function from the center and the corner of the BZ in the selected temperature intervals (Fig. 3a,b). Both plots confirm the previously detected in Fig. 2 trend. The measure of the temperature-induced shifts can be derived comparing pairs of EDCs taken at low and high temperatures (Fig. 3c, d). The difference between the blue and red arrows in Fig. 3c is 9.5 meV, and is 24 meV in Fig. 3d, i.e. well beyond the experimental errors. We plot the positions of the strongest EDC maximum from the center of the BZ (black symbols) in the full studied range of temperatures in Fig. 3e. As Fig. 3a implies, it is even possible to track the maximum of the hole-like band which crosses the Fermi level (blue symbols) in a limited temperature interval. This peak can be directly seen in red EDC in Fig. 3d at approximately -20 meV binding energy. From Fig. 3e it is seen that both bands are sensitive to the temperature, but below the nematic transition they appear to shift with different speeds. This happens because of an additional splitting between these two bands caused by nematicity [2].

The bottoms of the electron bands (Fig. 3d) are situated lower in energy than the tops of the hole bands (Fig. 3c) and because of the stronger scattering they are not clearly separated in the EDC's lineshape [2, 27]. In order to avoid a complicated fitting of the EDCs which usually requires many parameters because of energy-dependent self-energy, we approximate their temperature dependence by tracking the position of maximum of a single broad feature above the nematic transition (Fig. 3f) and by the mean value of the binding energies of all four peaks at 6K. A sketch of the actual temperature evolution of the band structure based on our previous experimental results [2] is shown in Fig. 3f with pink dashed lines.

Obviously, such significant variations in energy of the bands should result in changes of the size of the Fermi surfaces. Direct comparison of the Fermi surface maps in Fig.4 a-d is in line with all previous statements and confirms the enlargement of both Fermi surfaces upon warming up the sample. We note, that in the case of hole-like Fermi surface, which though looks larger on the map, one should be careful when analyzing its size quantitatively. Usually used for this procedure distance between the peaks of $E_F$-MDC can be inconclusive because at high temperatures this lineshape is influenced by two additional factors: the distance between such peaks becomes comparable to their width and the top of another hole-like band approaches the Fermi level and its spectral weight modifies the $E_F$-MDC. Actually, both these factors visually reduce the size of the hole pocket in Fig. 4d. For this reason, we will choose different measures to quantify the observed temperature variations in the following.

We schematically summarize the observed in this study changes in Fig. 4e, f, which represent low and high temperature respectively. Here top panels show band structure and bottom ones show the Fermi surfaces.

The blue/red shifts in FeSe tend to decrease with raising the temperature thus resulting to simultaneous growth of all Fermi surfaces.

Variations of the electronic structure with temperature have been detected earlier and in different materials. For instance, charge density wave bearing TaSe$_2$ exhibits pronounced T-dependent Fermi surface shape and its nesting. This behavior was attributed to the presence of the pseudogap, nonmonotonically varying with temperature [30, 31]. Also in IBS the temperature dependence was found in undoped and electron-doped 122 materials [10, 32]. Both studies reported the shrinking of the hole pocket and expansion of the electron one and thus considerable increase of charge carriers density upon warming the samples up. Interestingly, the Fermi surface of hole doped 122 material taken at room temperature did not show any noticeable departure from the low-temperature version [8]. We point out that the electronic structure of FeSe is simpler than that of 122 compounds because it is tetragonal at higher temperatures, does not undergo folding due to SDW phase at low temperatures and is less three-dimensional. Therefore the temperature dependent variations should be seen more clearly in FeSe.

Two theoretical approaches seem to explain the blue/red shifts in IBS at low temperatures. The first one is the electronic instability called Pomeranchuk effect [13, 15, 16, 18, 19]. The forward scattering triggers the distortion of the Fermi surface which preserves the point group symmetry of the crystal. The area of both electron and hole pockets increases (or decreases) so that the total charge density remains constant [13, 15]. The second approach is based on the renormalization of the bands by spin fluctuations via self-energy effects [12, 14]. The blue/red shifts and shrinking of the FS reported by ARPES are considered as a direct consequence of the coupling to a bosonic mode upon proper account of particle-hole asymmetry and the multiband character. It has been also suggested [28] that the blue/red shifts can be understood as a suppression of nearest-neighbor hopping due to spin/orbital orderings.

It is the reported in the present study temperature dependence of the electronic structure, which may help to distinguish between these two theoretical approaches. The S+- Pomeranchuk effect will result in a mean-field order parameter which is the relative shift of the electron bands in the corners and hole bands in the center of the BZ. As any mean-field order parameter it will diminish as temperature increases and will disappear at particular critical temperature. In the case of spin-fluctuation induced shifts the latter will decrease with temperature as well, but this behavior will be dictated by the softening of the spin-fluctuation spectrum and therefore will evolve more smoothly.

In Fig. 4 g, h we present two quantities which can be considered as the energy- and momentum-derived order parameters of the phenomenon. Fig. 4g. shows a temperature evolution of the energy separation between the top of the middle hole band and value which we use for characterization of the EDC from the near the corner of the BZ (Fig. 3e.). This parameter is given in units of fraction of this distance to the one obtained in LDA calculations [2]. To estimate this parameter for the low temperatures we used the mean values of energy positions of all 4 peaks extracted from the EDC measured at 6K. The monotonic and slightly superliniear increase with a signature of the nematic transition reflects the experimental observations discussed above. The second parameter (Fig. 4h) is obtained by relating the momentum width of the electron pocket (Fig. 4 a, b) to the width obtained in band structure calculations. Here the temperature dependence only slightly deviates from linear above ~200 K without any pronounced signatures related to nematic transition.

The behavior of both parameters does not seem to be in an immediate agreement with either of the theoretical approaches mentioned above. If the curve from Fig. 4g does resemble the behavior of the typical mean-field order parameter expected in the case of S+- Pomeranchuk effect, the momentum-related quantity does not. The observed quasilinear behavior is most likely not expected also when considering the coupling to spin-fluctuations. The most striking result is that in spite of rather high temperatures, both parameters still indicate a considerable departure from LDA, implying either an extremely high onset temperature of the Pomeranchuk instability or a very unusual temperature decay of the spin-fluctuation spectrum. In any case, our results call for a more thorough theoretical investigation aiming at quantitative explanation of the temperature relaxation.

In conclusion, in this study we demonstrated that band structure near the corner and the center of the BZ monotonically changes with temperature from 6 K to room temperature. This change reduces the size of the blue/red shifts and expands both parts of the Fermi surface. Both parameters characterizing the effect in terms of energy and momentum are hard to reconcile with existing theoretical approaches.

**Acknowledgements**
We are grateful to Rafael Fernandes, Andrey Chubukov, Dung-Hai Lee, Lara Benfatto, Laura Fanfarillo and Alexandr Yaresko for fruitful discussions. We thank Diamond Light Source for access to beamline I05 (proposal si11543-1) and Timur Kim, Moritz Hoesch for the help at the beamline. The work is supported by DFG grants No. BO1912/2-2 and BO1912/3-1.

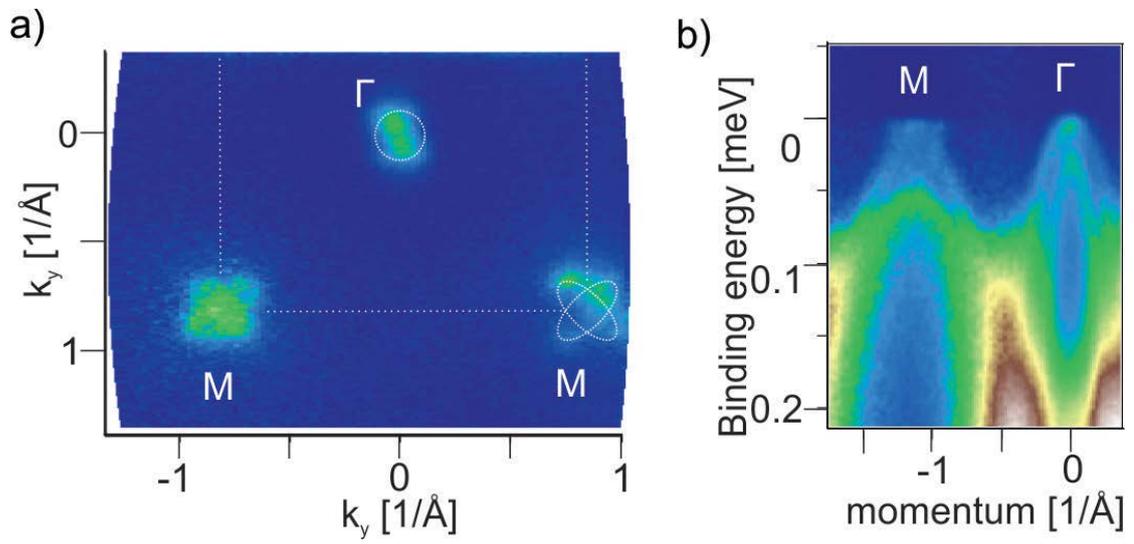

Fig. 1. a) ARPES-derived Fermi surface map of FeSe. Dashed lines schematically show the shape of the pockets in tetragonal phase and boundaries of the BZ; b) ARPES intensity along the diagonal of the BZ (Γ-M direction).

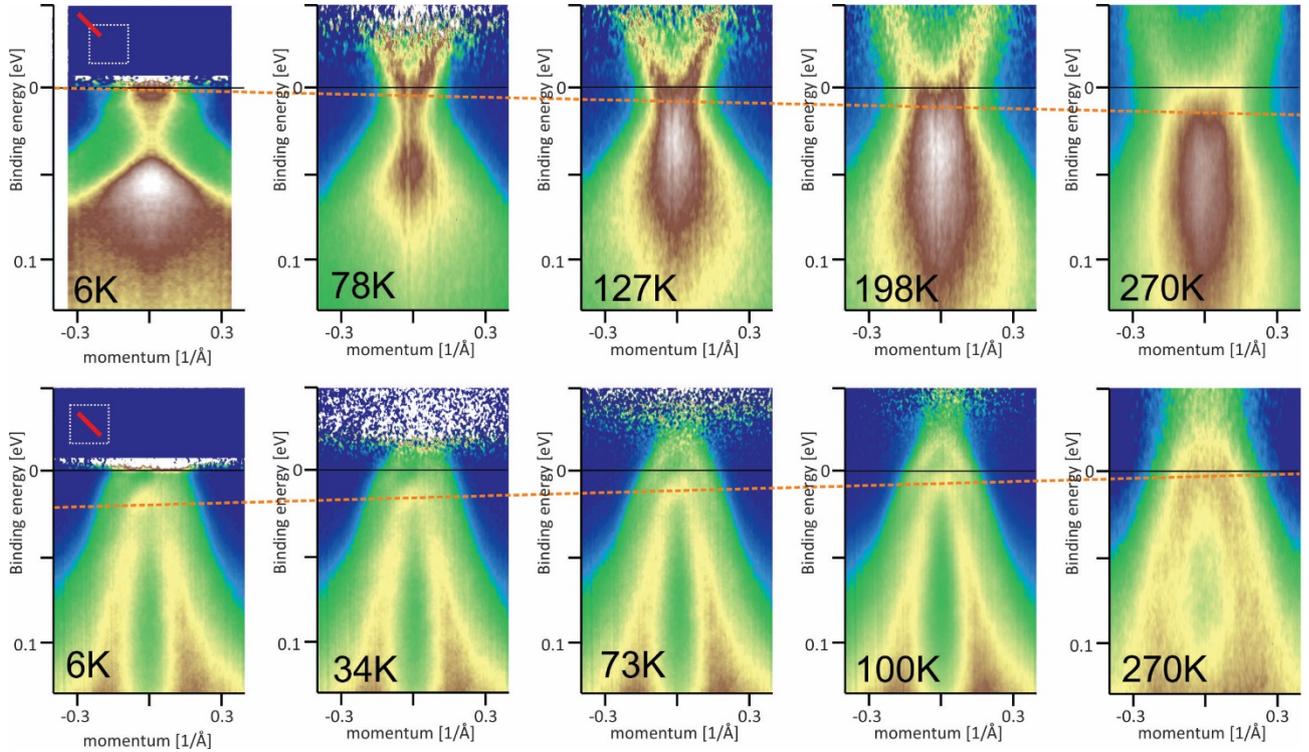

Fig. 2. ARPES intensity plots taken at different temperatures from 6 K to 270 K along the diagonal direction through the corner (upper panels) and the center (bottom panels) of the BZ.

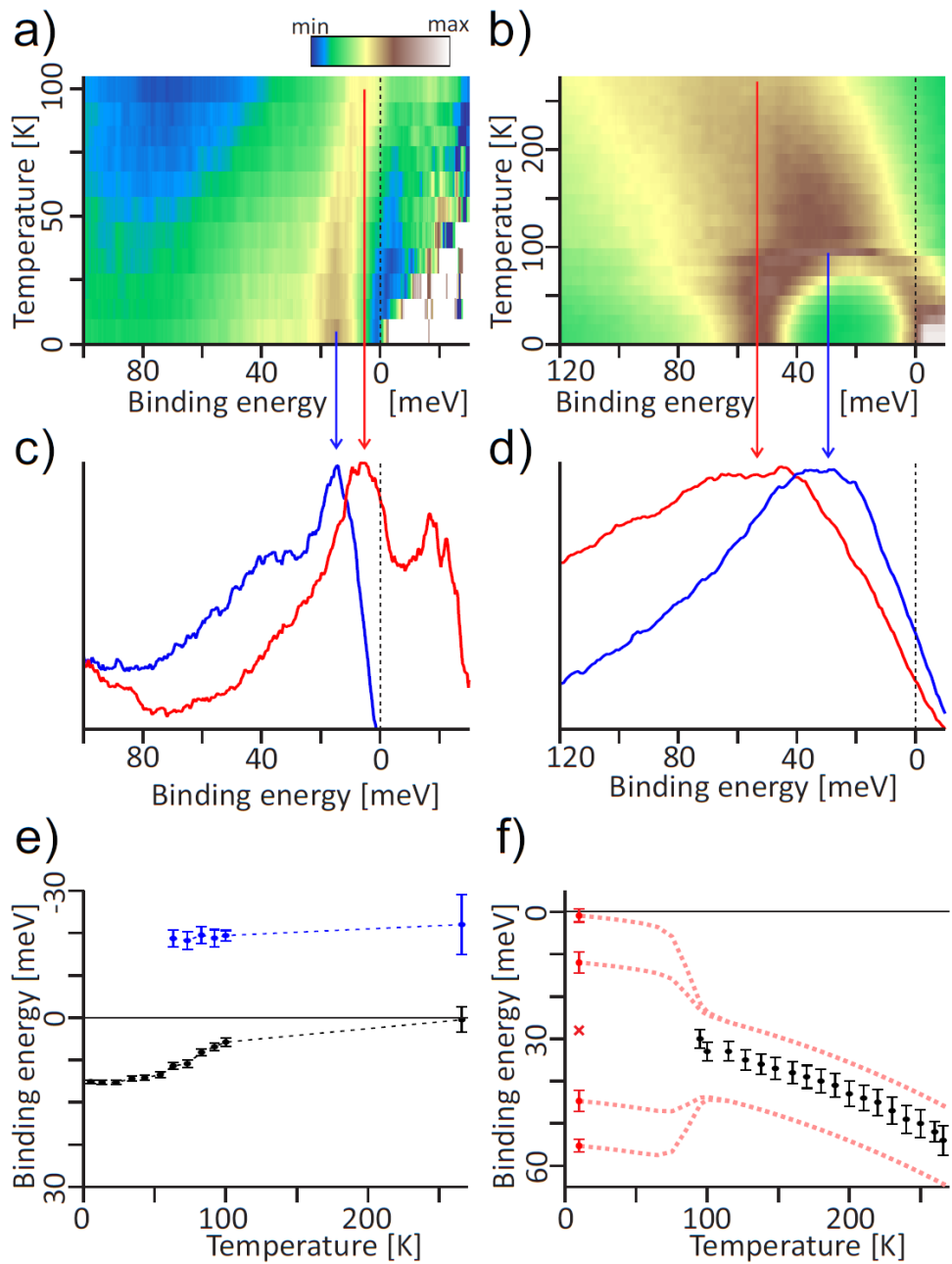

Fig. 3. a) Intensity plots for single EDCs from the center of the BZ as a function of temperature. All EDCs are divided by the Fermi function. b) the same for EDCs from the corner of the BZ; c), d) comparison pairs of EDCs with the highest and the lowest temperatures from panels a) and b) respectively; e) position of a maxima of EDCs from the panel a) (black dots), which represents top of lower and higher hole bands (black and blue dots respectively); f) position of 2 maxima of EDCs from the panel b., position of 4 peaks extracted from EDC measured for 6K (red dots) and their mean value (red cross); sketch of the temperature evolution of the band structure based on our experimental results (pink lines).

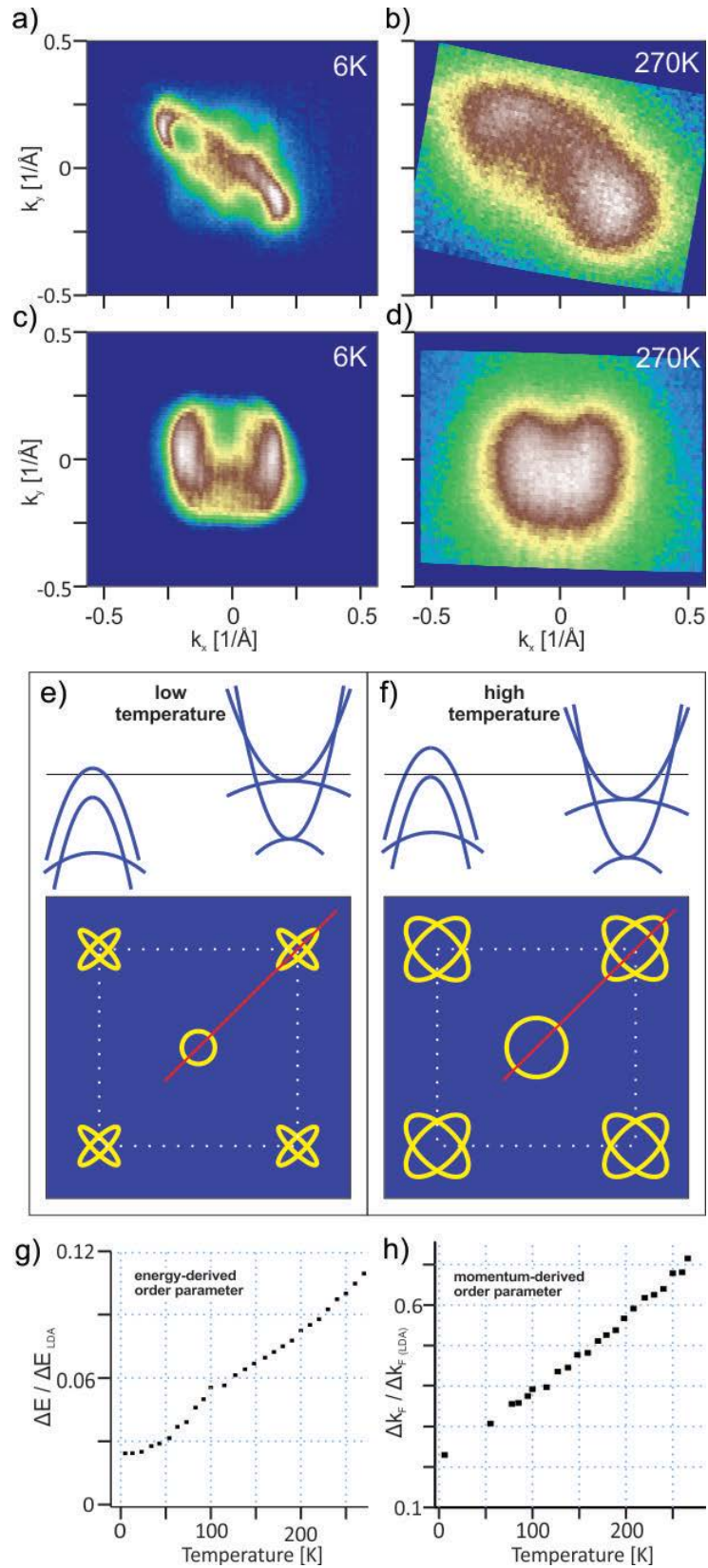

Fig. 4. a), c) Fermi surface maps measured at 6 K near the corner and the center of the BZ respectively; b), d) similar maps measured at 270 K. Sketches in plots e), f) represent band structure and Fermi surface at low and high temperature respectively. g) Energy distance between the top of the middle hole band and bottoms of electron pockets (see Fig. 3 e,f) normalized to calculated value. h) Momentum width of the electron pocket normalized to the calculated value.